\documentclass[aps,prc,floatfix,nofootinbib,showpacs,superscriptaddress,twocolumn]{revtex4-1}
\usepackage{amssymb}
\usepackage{amsmath}
\usepackage{graphics}
\usepackage{epsfig}
\usepackage[dvips]{color}
\usepackage{pzccal}

\newcommand{\be}{\begin{equation}}
\newcommand{\ee}{\end{equation}}
\newcommand{\bea}{\begin{eqnarray}}
\newcommand{\eea}{\end{eqnarray}}
\newcommand{\beas}{\begin{eqnarray*}}
\newcommand{\eeas}{\end{eqnarray*}}

\begin{document}
\title{Dynamical chiral symmetry breaking and the fermion--gauge-boson vertex}
\author{A.~Bashir}
\affiliation{Instituto de F\1sica y Matem\'aticas, Universidad Michoacana de San Nicol\'as de Hidalgo, Edificio C-3, Ciudad Universitaria, Morelia, Michoac\'an 58040, M\'exico}
\affiliation{Physics Division, Argonne National Laboratory, Argonne, Illinois 60439, USA}
\affiliation{Center for Nuclear Research, Department of Physics, Kent
State University, Kent OH 44242, USA}

\author{R.~Bermudez}
\affiliation{Instituto de F\1sica y Matem\'aticas, Universidad Michoacana de San Nicol\'as de Hidalgo, Edificio C-3, Ciudad Universitaria, Morelia, Michoac\'an 58040, M\'exico}

\author{L.~Chang}
\affiliation{Physics Division, Argonne National Laboratory, Argonne, Illinois 60439, USA}

\author{C.\,D.~Roberts}
\affiliation{Physics Division, Argonne National Laboratory, Argonne, Illinois 60439, USA}
\affiliation{Department of Physics, Illinois Institute of Technology, Chicago, Illinois 60616, USA}

\date{15 December 2011}

\begin{abstract}
We present a workable model for the fermion-photon vertex, which is expressed solely in terms of functions that appear in the fermion propagator and independent of the angle between the relative momenta, and does not explicitly depend on the covariant-gauge parameter.  It nevertheless produces a critical coupling for dynamical chiral symmetry breaking that is practically independent of the covariant-gauge parameter and an anomalous magnetic moment distribution for the dressed fermion that agrees in important respects with realistic numerical solutions of the inhomogeneous vector Bethe-Salpeter equation.
\end{abstract}


\maketitle

\section{Introduction}
\label{sec:introduction}
The last decade has seen a crystallisation of ideas regarding the nature of the dressed-gluon and -quark propagators in QCD.  In Landau gauge the dressed gluon two-point function is widely held to be described by a momentum-dependent mass function, $m_g^2(k^2)$.  Its magnitude is large at infrared momenta: $m_g^2(k^2 \sim 0) \simeq (2-4\,\Lambda_{\rm QCD})^2$.  However, it vanishes with increasing spacelike momenta: $m_g^2(k^2\gg \Lambda_{\rm QCD}^2) \sim 1/k^2$, thereby maintaining full accord with perturbative QCD.  Background and context for these observations may be found, e.g., in Refs.\,\cite{Bowman:2004jm,Aguilar:2010gm,Cucchieri:2011um,Boucaud:2011ug,Pennington:2011xs}, and citations therein and thereto.

Similarly, the dressed quark two-point function is described by two momentum-dependent functions: a wave-function renormalisation, $Z(p^2)$, and mass function, $M(p^2)$, both of which are strongly modified from their perturbative forms for $p^2 \lesssim (5\,\Lambda_{\rm QCD})^2$.  In fact, from the confluence of results on $M(p^2)$ obtained with Dyson-Schwinger equations (DSEs) in QCD and numerical simulations of the lattice-regularised theory, evident, e.g., in Refs.\,\cite{Bowman:2002bm,Maris:2002mt,Bhagwat:2003vw,Bowman:2005vx,Bhagwat:2006tu,Kamleh:2007ud}, a widespread appreciation has emerged of the reality and impact of dynamical chiral symmetry breaking (DCSB) in the strong interaction.

Since the two-point functions of elementary excitations are strongly modified in the infrared, one must accept that the same is generally true for three-point functions; i.e., the vertices.  This was actually realised early on, with studies of the fermion--gauge-boson vertex in Abelian gauge theories \cite{Ball:1980ay} that have inspired numerous ensuing analyses.  The importance of this dressing to the reliable computation of hadron physics observables was exposed in Refs.\,\cite{Frank:1994mf,Roberts:1994hh,Munczek:1994zz,Bender:1996bb}, insights from which have subsequently been exploited effectively; e.g., Refs.\,\cite{Maris:1997hd,Maris:2000sk,Eichmann:2008ef,Cloet:2008re,Chang:2010hb,Eichmann:2011vu,%
Chang:2011ei,Chang:2011vu,Wilson:2011rj}.

Many studies of hadron physics observables have employed an \emph{Ansatz} for the fermion--photon vertex.  The best are informed by analyses that emphasise the constraints of quantum field theory, amongst which are that the vertex should \cite{Ball:1980ay,Curtis:1990zs,Burden:1993gy,Dong:1994jr,Bashir:1994az,Bashir:2004mu,%
Kizilersu:2009kg,Bashir:2011vg}: be free of kinematic singularities; ensure gauge covariance and invariance; and assist in providing for the multiplicative renormalisability of solutions to the DSEs within which it appears.  \emph{Ans\"atze} that are largely consistent with these constraints have also been used to represent the dressed quark-gluon vertex.  In this connection, perhaps, it is clearest that such considerations are not sufficient to fully determine the vertex.  As an example, the Ball-Chiu vertex \cite{Ball:1980ay} augmented by the Curtis-Pennington extension \cite{Curtis:1990zs} is unable to explain the mass splitting between the $\rho$- and $a_1$-mesons, parity partners in the spectrum.  The minimum required to understand this is inclusion of a dressed-quark anomalous chromomagnetic moment \cite{Chang:2011ei}, the presence and strength of which are driven by DCSB \cite{Chang:2010hb,Kochelev:1996pv,Diakonov:2002fq} and confirmed by numerical simulations of lattice-QCD \cite{Skullerud:2003qu}.

We note that extending lattice analyses to the entire kinematic domain of spacelike momenta relevant to the numerous uses of the fermion--gauge-boson vertex is numerically challenging \cite{Kizilersu:2006et}.  We suspect that absent an appreciation within the lattice-QCD community of the physical importance of this problem, much time will elapse before new results are available.  This magnifies the importance of studies in the continuum.

It is in this context that we are motivated to readdress the task of constructing an efficacious and workable vertex \emph{Ansatz}.  Owing to sensible considerations regarding tractability, the most recent detailed studies \cite{Kizilersu:2009kg,Bashir:2011vg} have deliberately overlooked the role of DCSB in building the fermion--gauge-boson vertex.  Herein, informed by recent developments in hadron physics phenomenology \cite{Chang:2010hb,Chang:2011ei}, notably the dynamical generation of an anomalous magnetic moment for perturbatively massless fermions, we develop a practical \emph{Ansatz} for the fermion-photon vertex that produces a gauge-independent critical coupling for DCSB in QED and shows some promise as a tool for hadron physics phenomenology.

Section~\ref{sec:gapQED} provides some background on the coupling of a dressed-fermion to a photon.  Our \emph{Ansatz} is developed in Sec.\,\ref{sec:Ansatz} and employed to find a critical coupling for DCSB in Sec.\,\ref{sec:alphac}.  Section~\ref{sec:SandW} illustrates a phenomenological utility of the model and explains some obvious weaknesses, and Sec.\,\ref{sec:epilogue} is an epilogue.


%
\section{Gap Equation in QED}
\label{sec:gapQED}
Much of the progress toward understanding DCSB and the fermion--gauge-boson vertex has followed from studies of the gap equation, which in QED can be written
\begin{eqnarray}
\nonumber
S(p)^{-1} & = & Z_2(i\gamma\cdot p + m^{\rm bm}) \\
&+& Z_1 \!\!\int^\Lambda \!\frac{d^4 k}{4\pi^3}\,\alpha \Delta_{\mu\nu}(k-p) \gamma_\mu S(k) \Gamma_\nu(k,p)\,, \label{eq:gap}
\end{eqnarray}
where: we employ a Poincar\'e invariant regularisation of the integral, with $\Lambda$ the regularisation mass-scale, which typically doubles as the renormalisation point in DSE studies of QED; $Z_2(\Lambda)$ is the fermion wave-function renormalisation ($Z_1=Z_2$ in QED); $\alpha$ is the fine-structure constant; and the dressed-photon propagator is
\begin{equation}
\Delta_{\mu\nu}(q) = \left[\delta_{\mu\nu} - \frac{q_\mu q_\nu}{q^2}\right]\frac{1}{q^2[1+\Pi(q^2)]} + \xi \frac{q_\mu q_\nu}{q^4}\,,
\end{equation}
with $\xi$ the covariant-gauge parameter.  Whilst our vertex \emph{Ansatz} is always consistent with one-loop QED perturbation theory, aspects of the infrared behaviour we elucidate are particular to the quenched theory; viz., $\Pi(q^2)\equiv 0$.  We use a Euclidean metric:  $\{\gamma_\mu,\gamma_\nu\} = 2\delta_{\mu\nu}$; $\gamma_\mu^\dagger = \gamma_\mu$; $\sigma_{\mu\nu}=(i/2)[\gamma_\mu,\gamma_\nu]$; $a \cdot b = \sum_{i=1}^4 a_i b_i$; and $P_\mu$ spacelike $\Rightarrow$ $P^2>0$.

The solution of Eq.\,\eqref{eq:gap} has the form
\begin{equation}
S(p)
= \frac{1}{i \gamma \cdot p A(p^2)  + B(p^2)}
= \frac{Z(p^2)}{i \gamma \cdot p + M(p^2)}\,.
\end{equation}
In order to study DCSB, one must define a chiral limit and explore the behaviour of $M(p^2)$ as the value of the fine-structure constant is varied.  This is straightforward if the fine-structure constant is less than some critical value, denoted by $\alpha_c$, for then $m^{\rm bm}(\Lambda)\equiv 0$ defines the chiral limit and $M(p^2) \equiv 0$ is the only solution for the mass function.  It might, however, be viewed as problematic for $\alpha > \alpha_c$ because four-fermion operators become relevant in strong-coupling QED and must be included in order to obtain a well-defined continuum limit \cite{Rakow:1990jv,Reenders:1999bg}.  This complication is not of concern to us because one can obtain the critical coupling by approaching this value from below and at strong coupling one can view the cutoff version of the theory as an illustrative model.

Owing to the Ward-Takahashi identity:
\begin{equation}
(k-p)_\mu i \Gamma_\mu(k,p) = S^{-1}(k) - S^{-1}(p)
\end{equation}
(or the Slavnov-Taylor identity in non-Abelian theories), eleven independent tensor structures are required to fully express a fermion--gauge-boson vertex.  Furthermore, $\Gamma_\mu(k,p)$ can always be decomposed into two pieces:
\begin{equation}
\Gamma_\mu(k,p) = \Gamma^{\rm BC}_\mu(k,p)+\Gamma^{\rm T}_\mu(k,p)\,,
\end{equation}
with $(k-p)_\mu\Gamma^{\rm T}_\mu(k,p)=0$ and \cite{Ball:1980ay}
\begin{eqnarray}
\nonumber  
i\Gamma_\mu^{\rm BC}(k,p)  & = &
i\Sigma_A(k^2,p^2)\,\gamma_\mu + 2 \ell_\mu \left[ i\gamma\cdot \ell \,\Delta_A(k^2,p^2)  \right. \\
&&  \left. + \Delta_B(k^2,p^2)\right] , \label{bcvtx}\\
&:=& \sum_{i=1}^3 \lambda_i(k^2,p^2) \, i L^i_\mu(k,p) \,,
\label{bcvtx2}
\end{eqnarray}
for an Abelian theory,
where $2 \ell = k+p$,
\begin{subequations}
\begin{eqnarray}
\Sigma_{\phi}(k^2,p^2) &=& \frac{1}{2}[\phi(k^2)+\phi(p^2)]\,, \\
\Delta_{\phi}(k^2,p^2) &=& \frac{\phi(k^2)-\phi(p^2)}{k^2-p^2}\,,
\end{eqnarray}
\end{subequations}
$\lambda_1(k^2,p^2) = \Sigma_{A}(k^2,p^2)$, $\lambda_{2,3}(k^2,p^2) = \Delta_{A,B}(k^2,p^2)$.  We remark that some hints for a practical extension of Eq.\,\eqref{bcvtx} to QCD can be found in Ref.\,\cite{Bhagwat:2004kj} and it is conceivable that transverse symmetry transformations might assist in placing constraints on $\Gamma^{\rm T}_\mu(k,p)$ \cite{He:2009sj}.


%
\section{Vertex \emph{Ansatz}}
\label{sec:Ansatz}
Eight independent tensors are required in order to specify the transverse vertex:
\begin{equation}
\label{vertexgeneral}
\Gamma_\mu^T(k,p) = \sum_{j=1}^8 \tau^j(k^2,p^2,k\cdot p)\, T_\mu^j(k,p)\,.
\end{equation}
The following decomposition was introduced in Ref.\,\cite{Ball:1980ay}
\begin{subequations}
\label{TsEuclidean}
\begin{eqnarray}
T^1_{\mu} (k,p) &=& i \left[  p_\mu (k\cdot q) - k_\mu (p\cdot q) \right],\,q=k-p\,, \\
T^2_{\mu} (k,p) &=& -i T_{1\mu} (\gamma \cdot k + \gamma \cdot p)\,,\\
\label{T3mu}
T^3_{\mu} (k,p)  &=&  q^2 \gamma_\mu - q_\mu \gamma \cdot q =: q^2 \gamma_\mu^{\rm T}\,, \\
T^4_{\mu} (k,p)  &=&   iT_{1\mu} p_\nu k_\rho \sigma_{\nu\rho} \,,\\
T^5_{\mu} (k,p)  &=&  \sigma_{\mu\nu} q_\nu \,,\\
T^6_{\mu} (k,p)  &=&  -\gamma_\mu (k^2- p^2) + (k+p)_\mu \gamma \cdot  q \,, \\
\nonumber T^7_{\mu} (k,p)  &=&  \frac{i}{2} (k^2 - p^2) [ \gamma_\mu (\gamma \cdot  k + \gamma \cdot p) -(k+p)_\mu]\\
&& + (k+p)_\mu   p_\nu k_\rho \  \sigma _{\nu\rho} \,,\\
T^8_{\mu} (k,p)  &=&  
k_\mu \gamma \cdot p
- p_\mu \gamma \cdot  k
-i \gamma_\mu p_\nu k_\rho \sigma _{\nu\rho} \,,
\end{eqnarray}
\end{subequations}
and has since been used widely.  As we shall see, however, it has a couple of pitfalls.

A model for the vertex consists in a choice for the scalar-valued functions $\{\tau^j,j=1,\ldots,8\}$.  Following but expanding upon Ref.\,\cite{Bashir:2011vg}, we choose
\begin{subequations}
\label{ansatz}
\begin{eqnarray}
\label{tau1}
\tau _1(k^2,p^2) &=&  \frac{a_1\, \Delta_B(k^2,p^2)}{(k^2+p^2)} \,,\\
\label{tau2}
 \tau_2(k^2,p^2) &=& \frac{a_2\, \Delta_A(k^2,p^2)}{(k^2+p^2)} \,,\\
\tau_3(k^2,p^2) &=& a_3\, \Delta_A(k^2,p^2)\,, \\
\tau_4(k^2,p^2) &=&
\frac{a_4 \, \Delta_B(k^2,p^2)}{[k^2+M^2(k^2)[p^2+M^2(p^2)]} \,, \\
\tau_5(k^2,p^2) &=& a_5\, \Delta_B(k^2,p^2) \,,\\
\tau_6(k^2,p^2) &=&
 \frac{a_6(k^2+p^2) \, \Delta_A(k^2,p^2)}{[(k^2-p^2)^2+(M^2(k^2)+M^2(p^2))^2]}\,,\rule{2em}{0ex} \\
\tau_7(k^2,p^2) &=& \frac{a_7\, \Delta_B(k^2,p^2)}{(k^2+p^2)} \,,\\
\tau_8(k^2,p^2) &=&  a_8 \Delta_A(k^2,p^2)\,,
\end{eqnarray}
\end{subequations}
where $\{a_i,i=1,\ldots,8\}$ are momentum-independent constants.  This construction draws from a direct comparison with the structural dependence of the Ball-Chiu vertex on the functions that  constitute the fermion propagator; and the momentum-dependence of each term guarantees that from our \emph{Ansatz} one recovers a vertex which possesses the appropriate leading-order perturbative behaviour for $k^2\gg p^2$.  The coefficients $a_i$ are not independent.  As we now illustrate, they are interconnected by numerous constraints from perturbative QED and gauge covariance.


\subsection{One-loop Perturbation Theory}
\label{subsec:oneloop}
At one-loop in an arbitrary covariant gauge, the fermion-photon vertex obeys:
\begin{eqnarray}
\label{oneloopvertex}
\Gamma_{\mu}^{\rm T}(k,p) & \stackrel{k^2 \gg p^2}{=} &
- \frac{\alpha \xi}{8 \pi} \; {\rm ln}\frac{k^2}{p^2} \,
\left[ \gamma_{\mu} - \frac{k_{\mu} \gamma\cdot k}{k^2} \right]\,.
\end{eqnarray}
In the context of Eqs.\,\eqref{ansatz}, this demands \cite{Curtis:1990zs,Bashir:2000rv,Bashir:2004mu}
\begin{equation}
\label{a36}
a_3+a_6 = {1}/{2}\,.
\end{equation}
In addition, given the anticipated asymptotic behaviour of the scalar functions in the dressed-fermion propagator, then for $k^2\gg p^2$ the other terms in Eqs.\,\eqref{ansatz} decay as follows, up to $\ln [k^2/p^2]$-factors:
\begin{equation}
\tau _1 < \frac{1}{k ^ 2}, \;
\tau _{2,4,5,7} < \frac{1}{k ^ 3}, \;
\tau _8 < \frac{1}{k }; \label{tauotros}
\end{equation}
i.e., damping in agreement with one-loop perturbation theory \cite{Kizilersu:1995iz,Davydychev:2000rt}.  In building an \emph{Ansatz}, it is natural to insist on such correspondences with perturbation theory.

Some comments on Eq.\eqref{a36} are necessary.  Consider the choice $(a_3,a_6)=(0,1/2)$, which corresponds precisely to the \emph{Ansatz} of Ref.\,\cite{Curtis:1990zs}.  Taken as a statement about the vertex on a carefully defined domain of asymptotically-large spacelike momenta, this assumption is internally consistent.  However, whilst there are mitigating considerations, implemented as a constraint on the vertex for the entire domain of $(k^2,p^2,q^2)$, the assumption $a_3\equiv 0$ is generally mistaken.  The quantity $a_3$ is associated with $T_\mu^3$ in Eq.\,\eqref{T3mu} and hence contributes as follows to the complete vertex:
\begin{equation}
\gamma_\mu^{\rm T} \, q^2 \tau_3(k^2,p^2,q^2) = \gamma_\mu^{\rm T} \, q^2 \, a_3 \,\Delta_A(k^2,p^2)\,.
\end{equation}
Since $\gamma_\mu^{\rm T}$ is the leading tensor structure associated with a vector meson bound-state and given the existence of the $\rho$-meson, no realistic solution of the inhomogeneous Bethe-Salpeter equation for the fermion-photon vertex can produce a coefficient of $\gamma_\mu^{\rm T}$ that is identically zero.  Notwithstanding this, the choice $(a_3,a_6)=(0,1/2)$ is not worse than using $\Gamma_\mu^{\rm BC}$ alone.


\subsection{Multiplicative Renormalisability}
\label{sec:MR}
In the Wigner phase, multiplicative renormalisability of the fermion propagator requires that $Z(p^2) = (p^2)^\nu$, where $\nu$ is an anomalous dimension \cite{Curtis:1990zs,Dong:1994jr,Bashir:1995qr}.  It is multiplicative renormalisability that ensures the absence of overlapping divergences in the tower of DSEs.  In quenched-QED, one finds \cite{Bashir:2004hh}
\begin{equation}
\label{bashirnu}
\nu = \mathpzc{a} \xi - \frac{3}{2} \mathpzc{a}^2 + \frac{3}{2} \mathpzc{a}^3 + {\cal O}(\mathpzc{a}^4)\,,\; \mathpzc{a} = \frac{\alpha}{4\pi}\,.
\end{equation}

In connection with our vertex \emph{Ansatz}, power law behaviour of $Z(p^2)$ is guaranteed so long as
\begin{equation}
\label{a2368}
 1 + a_2 + 2 (a_3 - a_6 + a_8) =0 \,.
\end{equation}
This condition ensures the absence of $\ln\Lambda$-divergences and exposes the constraint
\begin{eqnarray}
f_1(\xi,\alpha,\nu;a_6)&:=& -\frac{4 \pi}{\alpha} + \frac{\xi}{\nu} +
 ( 1-2a_6) \times \left[ \frac{3}{2}+ \pi\cot\pi\nu \right.\nonumber \\
 &&  \left.   - \frac{1}{\nu + 2}
 - \frac{1}{\nu +1} - \frac{1}{\nu}  \right] =0 \,. \label{adnan1}
\end{eqnarray}
The choice $a_6=1/2$, discussed in connection with Eq.\,\eqref{oneloopvertex}, produces that anomalous dimension associated solely with gauge covariance \cite{Burden:1993gy,Dong:1994jr}.
%
This is the leading-order result.  It is known to be incomplete but corrections are higher-order in $\alpha$ and depend on the truncation, as evident via Eq.\,\eqref{bashirnu}.  They become crucial in strongly coupled theories \cite{Maris:1996zg,Bashir:2008fk,Bashir:2009fv}.

As one approaches the bifurcation point associated with the onset of DCSB from below, the behaviour of $Z(p^2)$ begins to influence the dressed-fermion mass function, which itself possesses power-law behaviour in the ultraviolet \cite{Gusynin:1989mc,Bashir:2011ij}.  In this instance, elimination of $\ln \Lambda$-divergences requires
\begin{equation}
\label{a457}
\frac{1}{2} a_4 + 2 a_5 + a_7 = 0\,.
\end{equation}
The tensor $T^5_\mu$ is that matrix structure associated directly with the Pauli form factor of an on-shell fermion.  One should therefore expect a realistic vertex \emph{Ansatz} to have $a_5\neq 0$.  As a consequence, at least one of $a_{4,7}$ must be nonzero.


\subsection{Anomalous Magnetic Moment}
\label{subsec:AMM}
Pursuing this point, consistency with the one-loop result for the anomalous magnetic moment of an on-shell fermion with mass $m$ \cite{Schwinger:1948iu} entails
\begin{equation}
\label{a2358}
  - m \lambda_2 + \lambda_3 + \tau_5 + m \tau_8 =
  \frac{\alpha}{4 \pi m}\,,
\end{equation}
where the functions are evaluated at $k^2=p^2=-m^2$, $q^2=0$.  In the context of our model, the content of Eq.\,\eqref{a2358} is readily illustrated: in Landau gauge ($\xi=0$) $\lambda_2 = 0 = \tau_8$ at one-loop order
and Eq.\,\eqref{a2358} becomes the constraint
\begin{equation}
(1+a_5) m B^\prime(-m^2) = \frac{\alpha}{4\pi}\,.
\end{equation}
Using the one-loop expression for $B(p^2)$, neglecting $\ln p^2/m^2$ terms for simplicity, then the constraint entails $a_5=(-4/3)$.  (Inclusion of the $\ln$-term reduces this value by 11\%, which is a negligible effect for our purposes.)

Observe now that the tensors $T^{1,2}_\mu$ can be reexpressed:
\begin{equation}
T_\mu^1 =  i q^2 \ell_\mu^{\rm T},\;
T_\mu^2 = \frac{1}{2} \gamma\cdot \ell \, T_\mu^1.
\end{equation}
Following numerous earlier studies, we have composed a vertex \emph{Ansatz} based on these tensors and required that their coefficient functions be free of kinematic singularities; viz., regular at $q^2=0$.  Using Eq.\,\eqref{ansatz}, their net contribution to the vertex is
\begin{equation}
\frac{q^2 \, a_1 \Delta_B }{2 \ell^2 + \frac{1}{2} q^2} i \ell_\mu^{\rm T}
+ \frac{q^2 \, a_2  \Delta_A }{2 \ell^2 + \frac{1}{2}q^2} \frac{1}{2} \gamma\cdot\ell\, \ell_\mu^{\rm T}\,.
\end{equation}
The asymptotic domain $k^2\gg p^2$ corresponds to large values of $\ell^2 \sim q^2/4$, upon which the contribution is therefore
\begin{equation}
\label{alternativetensor12}
a_1 \Delta_B  i \ell_\mu^{\rm T}
+ a_2 \Delta_A \frac{1}{2} \gamma\cdot\ell\, \ell_\mu^{\rm T}\,.
\end{equation}
Plainly, the kinematic dependence on $q^2$ plays no role asymptotically and the standard analysis proceeds without reference to it.  On the infrared domain, however, the kinematic dependence on $q^2$ is crucial.  With the choices described above, the contribution from $\tau_{1,2}$ to the fermion's anomalous magnetic moment vanishes.  This is a pitfall of the basis in Eq.\,\eqref{TsEuclidean}.

With hindsight, one could equally have chosen to use the tensor basis in Ref.\,\cite{Chang:2010hb}, which corresponds to
\begin{equation}
\tilde T_\mu^1 = -i \ell_\mu^{\rm T},\;
\tilde T_2^\mu = -i \gamma\cdot \ell \tilde T_\mu^1.
\end{equation}
In this case Eqs.\,\eqref{tau1}, \eqref{tau2} suggest alternative natural choices for the coefficient functions:
$\tilde\tau_1 = a_1 \Delta_B$, $\tilde\tau_2 = (a_2/2) \Delta_A$;
Eq.\,\eqref{alternativetensor12} represents the vertex contribution for all values of $q^2$; and the fermion's anomalous magnetic moment receives contributions from both these terms.  Equation~\eqref{a2358} is then modified as follows:
\begin{equation}
\label{a235812}
  - m \lambda_2 +  \lambda_3  + \tilde \tau_1 + m \tilde \tau_2 +
  \tau_5 + m \tau_8 =
  \frac{\alpha}{4 \pi m}\,.
\end{equation}
A comparison with leading-order perturbation theory would now inform constraints on additional structures in the vertex.  For example, Eq.\,\eqref{a235812} and  Ref.\,\cite{Davydychev:2000rt} together entail that $\tilde\tau_{1,2}(k^2,k^2)$ should assume functional forms that vanish at leading-order in Landau gauge.

At this point a comparison with Ref.\,\cite{Maris:1999bh} is useful.  The function $q^2\tau_1(q^2)$ here corresponds to $F_5(q^2)$ therein, which does not vanish at $q^2=0$.  Similarly, $\hat F_{5,7}(q^2)$ in Ref.\,\cite{Chang:2010hb}, which correspond to $q^2\tau_{1,2}(q^2)$ herein, are nonzero and significant at $q^2=0$.
This suggests that a tensor basis which avoids multiplicative factors of $q^2$ is better suited to building vertex \emph{Ans\"atze} intended for use on the entire domain of $(k^2,p^2,q^2)$ that is sampled in both nonperturbative solutions of truncated DSEs and applications in hadron physics.


\section{Critical Coupling for DCSB}
\label{sec:alphac}
Having detailed an \emph{Ansatz} and a number of constraints, we are now in a position to solve the gap equation; i.e., the coupled equations for $Z(p^2)$, $M(p^2)$.  These equations simplify in the neighbourhood of $\alpha_c$; viz., the coupling whereat a $M(p^2)\not\equiv 0$ solution bifurcates away from the $M(p^2)\equiv 0$ solution, which alone is possible in perturbation theory \cite{Atkinson:1986aw,Atkinson:1992tx,Atkinson:1993mz}.  The behaviour of the solutions near the bifurcation point may be investigated by performing functional differentiation of the gap equations with respect to $M(p^2)$ and evaluating the results at $M(p^2)=0$.  Practically, this amounts to analysing linearised forms of the original gap equations; i.e., the equations obtained by eliminating all terms of quadratic or higher order in $M(p^2)$.

Since the only mass-scale at the bifurcation point is the regularisation parameter, then in the vicinity of $\alpha_c$ one may uniformly approximate the mass function as $M(p^2) \sim (1/p^2)^{s}$, $s=1-\gamma_m/2$ with $\gamma_m$ the mass anomalous dimension.
Combined with our vertex \emph{Ansatz}, power-law behaviour for both $Z$, $M$ enables one to evaluate the angular and radial integrals in the gap equation.  This produced Eq.\,\eqref{adnan1} and here yields the following equation in the quenched truncation:
\begin{widetext}
\begin{eqnarray}
\label{adnan2}
\lefteqn{f_2(\xi,\alpha,\nu,s;a_1,a_2,a_4,a_6)= 0\,,}\\
\nonumber
 &=& -\xi + \frac{3}{2} \frac{\nu(\nu -s+1)}{1-s} \Bigg\{  -\pi\cot\pi\nu - \pi\cot\pi (s-\nu)  + \frac{1}{\nu} + \frac{1}{\nu +1} +\frac{1}{s}  + \frac{1}{1-s} +\frac{2}{s-\nu}  \\
\nonumber && + \left( 1 + \frac{a_2}{3}+2 a_6  \right) \left[ \pi\cot\pi s -\pi\cot\pi(s-\nu) \right]  - \left( \frac{a_2}{3} -2a_6 \right)
 \left[ \frac{1}{s-\nu} +\frac{1}{s-\nu -1} -\frac{1}{s}+\frac{1}{1-s}   \right]\\
\nonumber &&  - \frac{2a_2}{3} \Bigg[\frac{1}{s} -\frac{1}{s-\nu} - \overline{\psi}\left( \frac{s}{2} \right)+\overline{\psi} (s)  +\overline{\psi}\left( \frac{\nu -s}{2} \right) \nonumber - \overline{\psi}(\nu - s)  \Bigg]  \\
\nonumber && -  \frac{a_4}{12} \left[ \frac{1}{1 + s}
 + \frac{1}{2 - s} + \frac{1}{\nu - 1} - \frac{1}{\nu + 2} \right]
 + \left( \frac{a_7}{6} - \frac{a_1}{3} \right)
 \Bigg[ \frac{1}{2} \Bigg( \frac{3}{s} + \frac{1}{1 - s} + \frac{3}{\nu}
  - \frac{1}{\nu +1} - \pi \cot\pi s  \\
 \nonumber &&   - \pi \cot\pi \nu +
  \pi \cot\frac{\pi s}{2}
 + \pi \cot\frac{\pi \nu}{2} \Bigg)
 - \psi\left(-\frac{s}{2}\right)
 + \psi\left(\frac{\nu}{2}\right)
 + \overline{\psi}(s) - \frac{1}{2}
 \overline{\psi}\left(\frac{s}{2}\right) - \overline{\psi}(\nu) + \frac{1}{2} \overline{\psi}\left( \frac{\nu}{2} \right)
 \Bigg]  \\
&&  +  a_7 \Bigg[ \frac{1}{4} \left( \frac{5}{s} + \frac{3}{1 - s} + \frac{5}{\nu} -
       \frac{3}{\nu + 1} - \pi \cot\pi s - \pi \cot \pi \nu] \right)
 +   \psi(-s) - \psi\left(-\frac{s}{2}\right) - \psi(\nu) +
     \psi\left( \frac{\nu}{2} \right) \Bigg]
 \Bigg\} \,,
\end{eqnarray}
\end{widetext}
where: $\Gamma(x)$ is the Euler function; $\psi(x)$, its logarithmic derivative (digamma function); and $\overline{\psi} (x) =\psi(x) + \psi(-x)$.  N.B.\ The absence of power-law divergences is guaranteed so long as $\nu\in (-2,1)$, $s\in(0,1)$.  In addition, Eq.\,\eqref{a457} means that $a_7$ is not an independent parameter.

Our initial goal is to locate the bifurcation point, $\alpha_c$.
Consider Eqs.\,\eqref{adnan1}, \eqref{adnan2}.  One finds that for given values of $(\xi,\alpha)$ and vertex \emph{Ansatz} parameters, Eq.\,\eqref{adnan1} produces a solution for $\nu$.  Using these parameters in Eq.\,\eqref{adnan2}, one finds no solutions for $s$ if $\alpha<\alpha_c$ and two solutions if $\alpha>\alpha_c$: $f_2(\xi,\alpha,\nu,s;a_1,a_2,a_4,a_6)$ has the appearance of a flattened catenary; and the bifurcation point is found when the locations of this function's roots and its interior extremum coincide.  This condition may be imposed by requiring
\begin{equation}
f_3(\xi,\alpha,\nu,s;\vec{a}):= \frac{\partial }{\partial s} f_2(\xi,\alpha,\nu,s;\vec{a}) =0
\end{equation}
in addition to Eqs.\,\eqref{adnan1}, \eqref{adnan2}.

We now demand a little more; namely, that our \emph{Ansatz} produce a Landau-gauge value of $\alpha_c$ which agrees with that produced by the Ball-Chiu \emph{Ansatz} \cite{Ball:1980ay} and minimises $\partial\alpha_c/\partial\xi$ on $\xi\in[0,10]$.  (We choose the Ball-Chiu result to define $\alpha_c(\xi=0)$ because this \emph{Ansatz} is the minimal vertex consistent with the Ward and Ward-Takahashi identities.)  For this purpose $\alpha_c= \alpha_c(\xi;a_1,a_2,a_4,a_6)$; i.e., a function that describes a smooth surface in six dimensions, and our demand is straightforwardly mapped into a multidimensional extremisation problem.  Namely, find that set ${\cal E}=\{\check{a}_1,\check{a}_2,\check{a}_4,\check{a}_6\}$ which produces ${\rm min}\{(\partial\alpha_c(\xi;a_1,a_2,a_4,a_6)/\partial\xi)^2,\xi\in[0,10]\}$.  We obtain a solution with
\begin{equation}
\label{finalAnsatz}
\begin{array}{cccr|r|crr}
\check{a}_1 & \check{a}_2 & \check{a}_4 & \check{a}_6 & \check{a}_5 & \check{a}_3 & \check{a}_7 & \check{a}_8\\
0    & 3.4    &  6 &  -\frac{1}{2}   & -\frac{4}{3} &  1 & -\frac{1}{3} & -3.7
\end{array}\,,
\end{equation}
where $\check{a}_5$ was fixed following Eq.\,\eqref{a2358}, and $\check{a}_{3,7,8}$ are determined from the other \emph{Ansatz} parameters via Eqs.\,\eqref{a36}, \eqref{a2368}, \eqref{a457}.  N.B.\ With $a_1=0$, we comply with an observation made after Eq.\,\eqref{a235812}.

\begin{figure}[t]
\centerline{\epsfig{file=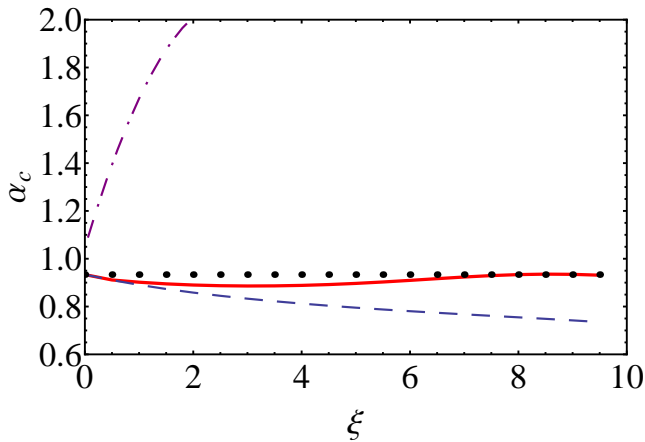, width=1\columnwidth}}
\caption{\label{fig:alphac}
Critical coupling for DCSB as a function of the gauge parameter, $\xi$: $\alpha_c(0)=0.934$.
\emph{Solid curve} -- Result obtained with the vertex \emph{Ansatz} defined by Eq.\,\protect\eqref{finalAnsatz}, which incorporates all structures that contribute to a fermion anomalous magnetic moment.
\emph{Circles} -- Allowing $a_1$ to depend on $\xi$, it is possible to ensure $\partial \alpha_c/\partial \xi \equiv 0$.
\emph{Dashed curve} -- Result obtained with the Curtis-Pennington vertex Ansatz \protect\cite{Curtis:1990zs}; and \emph{dash-dot curve} -- bare vertex result.}
\end{figure}

Our result is the solid curve in Fig.\,\ref{fig:alphac}, which illustrates that it is  straightforward to obtain a critical coupling that is almost insensitive to $\xi$ if all amplitudes are retained in the vertex.  Notably, this \emph{Ansatz} depends neither on $\xi$ nor the angle defined by $k\cdot p\,$.  Moreover, if any one of the parameters $a_i$ in Eq.\,\eqref{ansatz} is allowed to depend on $\xi$, then $\partial \alpha_c/\partial \xi \equiv 0$ is guaranteed.

It is important to observe that $\gamma_m=1.058$ in Landau gauge.  This result emphasises that four-fermion operators become relevant to QED for $\alpha>\alpha_c$ because the operator $(\bar\psi \psi)^2$ has dimension $2\times(3-\gamma_m)$.
As a consequence, one cannot sensibly compute a fermion condensate unless this operator is included.  That, however, introduces a new parameter, the operator's coupling strength, which cannot readily be constrained.  In these circumstances, no attempt to enforce $\xi$-independence of the condensate can produce additional meaningful constraints on the vertex.

It is worth digressing here and reiterating that Landau gauge occupies a special place \cite{Bashir:2008fk,Bashir:2009fv}.  For example, it is a fixed point of the renormalisation group; and that gauge for which the one-loop contribution to $A(p)$ vanishes in any number of dimensions in any renormalisable gauge theory (see Eq.\,\eqref{bashirnu} and Ref.\,\cite{Davydychev:2000rt}).  It follows from the latter that in Landau gauge any sensitivity to model-dependent differences between \emph{Ans\"atze} for the fermion-photon vertex are least noticeable.  One may therefore argue that all discussion of the gauge-parameter dependence of a vertex \emph{Ansatz} is moot, since the vertex in anything other than Landau gauge should simply be defined as the Landau-Khalatnikov-Fradkin (LKF) transform \cite{Landau:1955zz,Fradkin:1955jr,Johnson:1959zz,Zumino:1959wt} of the Landau gauge form.  The sensible implementation of this procedure guarantees gauge covariance and hence obviates any question about the gauge dependence of gauge invariant quantities.
Notwithstanding these observations, the LKF transform of a general vertex \emph{Ansatz} is practically difficult to obtain \cite{Burden:1993gy}, and, as we have illustrated herein, requiring gauge-parameter independence of physical quantities computed with a given vertex can inform and constrain the construction of an \emph{Ansatz}.

\begin{figure}[t] 
\centerline{\epsfig{file=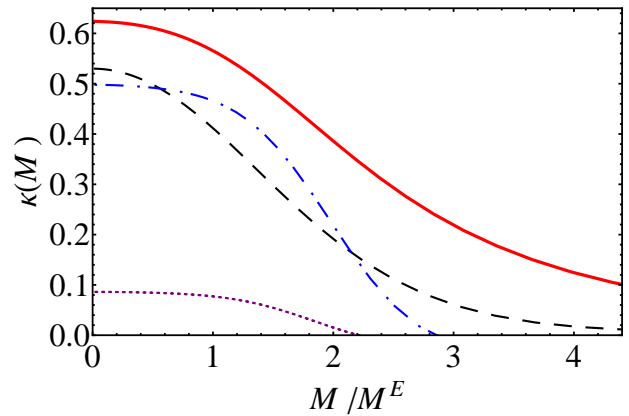, width=1\columnwidth}}
\caption{\label{fig:Dist}
Dynamically generated anomalous magnetic moment distribution for a perturbatively massless fermion.
\emph{Solid curve} -- computed from the interaction model described in Ref.\,\protect\cite{Qin:2011dd} using the vertex \emph{Ansatz} described herein.
\emph{Dashed curve} -- computed from the interaction model in Ref.\,\protect\cite{Chang:2010hb} using our vertex \emph{Ansatz};
and \emph{dot-dashed curve} -- distribution computed in Ref.\,\protect\cite{Chang:2010hb}, obtained via a symmetry-preserving simultaneous solution of the beyond-rainbow-ladder gap equation and inhomogeneous vector Bethe-Salpeter equation.
Between the last two curves, the mismatch at ultraviolet momenta is readily understood: our vertex \emph{Ansatz} expresses constraints from one-loop perturbation theory in a theory with a power-law fermion-fermion interaction, whereas the interaction in Ref.\protect\cite{Chang:2010hb} dies  exponentially.
\emph{Dotted curve} -- For comparison with the \emph{dashed} and \emph{dot-dashed} curves, the distribution obtained with any one of the \emph{Ans\"atze} in Refs.\,\protect\cite{Ball:1980ay,Curtis:1990zs,Kizilersu:2009kg}.}
\end{figure}


\section{Model's Strengths and Weaknesses}
\label{sec:SandW}
We now address aspects of an additional issue; namely, whether the \emph{Ansatz} expressed in Eqs.\,\eqref{vertexgeneral}--\eqref{ansatz}, \eqref{finalAnsatz} is adequate for use in nonperturbative studies of truncated DSEs or to describe the dressed-quark--photon coupling in hadron physics phenomenology.  Both applications sample domains of $(k^2,p^2,q^2)$ that stretch far outside those considered in asymptotic analyses.  We limit ourselves to the context provided by the anomalous magnetic moment of a dressed fermion, in part because above we have highlighted DCSB and this emergent phenomenon produces an anomalous magnetic moment for a perturbatively massless fermion: it is impossible for a truly massless fermion to possess a measurable anomalous magnetic moment \cite{Chang:2010hb}.

We follow Ref.\,\cite{Chang:2010hb} in constructing an anomalous magnetic moment distribution.  At each value of spacelike-$p^2$, define spinors to satisfy
\begin{equation}
\bar u(p,\mathpzc{M}) (i\gamma\cdot p + \mathpzc{M} ) = 0 = (i\gamma\cdot p + \mathpzc{M} )u(p,\mathpzc{M})\,,
\end{equation}
where $\mathpzc{M} = M(p^2)$; i.e., the mass-function evaluated at that value of $p^2$.  In this case
\begin{eqnarray}
\nonumber
\lefteqn{\bar u(k,\mathpzc{M}) \Gamma_\mu(k,p) u(p,\mathpzc{M})} \\
& = & \bar u(k,\mathpzc{M} )
\left[ F_1(q^2) + \frac{1}{2 \mathpzc{M}} \sigma_{\mu\nu} q_\nu F_2(q^2) \right]u(p,\mathpzc{M}) \,, \rule{1em}{0ex}
\end{eqnarray}
owing to a Gordon identity, and the fermion's anomalous magnetic moment is $\kappa(\mathpzc{M})=F_2(q^2=0,\mathpzc{M})/F_1(0,\mathpzc{M})$.  With the \emph{Ansatz} specified by Eqs.\,\eqref{vertexgeneral}--\eqref{ansatz}, \eqref{finalAnsatz},
\begin{equation}
\kappa(\mathpzc{M}) =
2 \mathpzc{M} \frac{(a_5 -1)  \delta_B + (1-a_8) \mathpzc{M} \delta_A}
    {\sigma_A - 2 \mathpzc{M}^2 \delta_A + 2 \mathpzc{M} \delta_B}\,,
\end{equation}
where $\sigma_A = \Sigma_A(\mathpzc{M}^2,\mathpzc{M}^2)$, $\delta_{A,B} = \Delta_{A,B}(\mathpzc{M}^2,\mathpzc{M}^2)$.  (N.B.\ The tensor denominated $T_\mu^8$ herein is associated with $\tau_4$ in Ref.\,\cite{Chang:2010hb}.)

The distribution is plotted in Fig.\,\ref{fig:Dist}, where the Euclidean constituent fermion mass $M^E = \{ p\,|p>0,p^2=M^2(p^2)\}$.  It is immediately apparent that DCSB produces a distribution that is large on the nonperturbative infrared domain: $\kappa_{M^E} = 0.45-0.55$, but vanishes with the strength of this dynamical effect.  More striking in the present context, however is the degree of similarity between the dashed and dot-dashed curves.  This signals that our vertex \emph{Ansatz} passes a nontrivial test.  Namely, whilst expressed solely in terms of the functions in the dressed-fermion propagator, it nevertheless produces a magnetic moment distribution in fair agreement with the most realistic symmetry-preserving solution of the inhomogeneous vector Bethe-Salpeter equation that is currently available.  It is therefore worth employing this Ansatz more widely; e.g., in the computation of hadron electromagnetic form factors \cite{Eichmann:2008ef,Cloet:2008re,Eichmann:2011vu,Wilson:2011rj}.

Notwithstanding this, there are some caveats that should be borne in mind.
As observed in Sec.\,\ref{subsec:oneloop}, no \emph{Ansatz} for the transverse part of the fermion-photon vertex is completely satisfactory if it does not express what might be called leakage into the spacelike region of spectral strength from both the two-pion continuum and the region of the $\rho$-meson pole.  If one concentrates on the spacelike domain, then such effects are maximal near $q^2=0$ and may be characterised by observing that they alter charge radii by $\lesssim 20$\% \cite{Alkofer:1993gu,Maris:1999bh,Roberts:2000aa,Roberts:2011wy,Wilson:2011rj}.
Related, and more important perhaps, are the limitations of the tensor basis in Eqs.\,\eqref{TsEuclidean} that we discussed in Sec.\,\ref{subsec:AMM}.  Namely, analyses concerned with domains of ultraviolet momenta are not necessarily a fair guide to the $q^2$-dependence of the vertex in the infrared.  Explicit computations \cite{Maris:1999bh,Chang:2010hb} present conflicts with expectations fed by \emph{Ans\"atze} built upon Eqs.\,\eqref{vertexgeneral}, \eqref{TsEuclidean}.
Finally, it is worth investigating how our \emph{Ansatz} fares in the DSE for the photon vacuum polarisation, which is known to be sensitive to features and kinematic domains that are not strongly constrained by the gap equation \cite{Kizilersu:2011zza}.  Such analyses would likely lead to a refinement of our model.


\section{Epilogue}
\label{sec:epilogue}
Motivated by the fact that knowledge of the dressed-fermion--gauge-boson vertex is critical to any continuum study of a gauge field theory, we have detailed a workable model for the dressed-fermion-photon vertex, $\Gamma_\mu(k,p)$.  It is expressed solely in terms of functions which appear in the dressed-fermion propagator, is independent of the angle defined by $k\cdot p\,$, and does not explicitly depend on the covariant-gauge parameter.  The \emph{Ansatz} is nevertheless consistent with constraints that have long been held important, namely: it is free of kinematic singularities; ensures gauge covariance and invariance in the application tested; and assists in providing for the multiplicative renormalisability of solutions to the DSEs within which it appears.

Significantly, the \emph{Ansatz} contains nontrivial factors associated with those tensors that are even in the number of Dirac matrices; i.e., whose appearance is expressly driven by dynamical chiral symmetry breaking in a perturbatively massless theory.  This novel feature enables a direct and positive comparison with the best available symmetry-preserving solutions of the inhomogeneous Bethe-Salpeter equation for the vector vertex.  The positive outcome indicates that our model might provide a much needed tool for use in Poincar\'e-covariant symmetry-preserving studies of hadron electromagnetic form factors.  Furthermore, given the general nature of our constraints and the simplicity of our construction, there is some room to hope that a straightforward extension of our approach might yield an \emph{Ansatz} adequate to the task of representing the dressed-quark-gluon vertex.


\section*{Acknowledgments}
We are grateful for useful input from C.~Chen, \mbox{S.-x.~Qin}, P.\,C.~Tandy and D.\,J.~Wilson.
This work was supported by the {\em Programa de Cooperaci\'on Bilateral M\'exico-Estados Unidos} (CONACyT 2009-2011) with counterpart funding from the U.\,S.\ National Science Foundation, under grant no.\ NSF-PHY-0903991; CONACyT project 46614-F;
Coordinaci\'on de la Investigaci\'on Cient\1fica (CIC) project no.\ 4.10;
%
%
and the U.\,S.\ Department of Energy, Office of Nuclear Physics, contract no.\ DE-AC02-06CH11357.


\end{document}